\documentclass[%
reprint,
superscriptaddress,
amsmath,amssymb,
aps,
prapplied,
]{revtex4-1}

\usepackage{graphicx}
\usepackage{dcolumn}
\usepackage{bm}
\usepackage[colorlinks, linkcolor=red, anchorcolor=blue,urlcolor=blue, citecolor=blue]{hyperref}
\usepackage{subfigure}

\newcommand{\beq}{\begin{equation}}
\newcommand{\eeq}{\end{equation}}
\newcommand{\beqa}{\begin{eqnarray}}
\newcommand{\eeqa}{\end{eqnarray}}

\begin{document}

\title{Inverse engineering and composite pulses for magnetization reversal}%

\author{Ze-Guo Song}
\affiliation{Department of Physics, Shanghai University, 200444 Shanghai, People's Republic of China}%

\author{Han Wu}
\affiliation{Department of Physics, Shanghai University, 200444 Shanghai, People's Republic of China}%

\author{Si Wang}
\affiliation{Department of Physics, Shanghai University, 200444 Shanghai, People's Republic of China}%

\author{Yue Ban}
\email[]{yban@shu.edu.cn}
\affiliation{Department of Electronic Information Materials, Shanghai University, 200444 Shanghai, People's Republic of China}%

\author{Xi Chen}
\email[]{xchen@shu.edu.cn}
\affiliation{Department of Physics, Shanghai University, 200444 Shanghai, People's Republic of China}%

\date{\today}

\begin{abstract}
We put forward a method for achieving fast and robust for magnetization reversal in a nanomagnet, by combining the inverse engineering and composite pulses. The magnetic
fields, generated by microwave with time-dependent frequency, are first designed inversely within short operation time, and composite pulses are further
incorporated to improve the fidelity through reducing the effect of magnetic anisotropy. The high-fidelity magnetization reversals are illustrated with numerical examples, and visualized on Bloch sphere.
The influence of damping parameters, relevant to the pulse sequence, is finally discussed based on Landau-Lifshitz-Gilber equation. These results
pave the way for precise but fast magnetization reversal or switching, with the applications in high density information storage and processing.

\end{abstract}

\maketitle

\section{INTRODUCTION}

The efficient initialization, manipulation, and readout
of electron spins are requisite in the field of spintronics and quantum information for the
implementation of high-density information storage and information processing
with a single electron spin qubits \cite{BookLoss,Loss}. Normally, in electric dipole spin resonance (EDSR) \cite{Poole}, microwaves drive an
electron to oscillate in the magnetic field or electric fields through spin-orbit coupling, producing a coherent spin manipulation \cite{Nowack,Tarucha,JQYouPRL}.
Besides, the Landau-Zener adiabatic schemes \cite{Landau} provides an analytical model for (effective) two-level quantum systems,
achieving the coherent manipulation of spin state \cite{EPL,Palli,Stepanenko}. However, such adiabatic processes and their variants require long operation time,
which becomes inefficient when the damping is considered under decoherent environment. To remedy it,
the inverse engineering \cite{Ruschhaupt,YuePRL,SarmaPRL,Shore} and quantum transitionless algorithm \cite{Berry,PRL105}, sharing the concept of shortcuts to adiabaticity (STA) \cite{STA},
and other relevant methods including optimal control \cite{Tannor,Boscain,Kontz} and composite pulses \cite{Levitt,Torosov-PRL}, have been proposed, which minic the adiabatic control
but in accelerated and robust ways.

In ferromagnetic nanostructures, rapid and robust magnetization reversal is of interest for both fundamental
physics and applications \cite{Bertotti,Sukhov,Barros11,Barros13,Cai,CaiPRX,Taniguchi,Klughertz}. A large number of experiments have illustrated that the microwave filed with appropriate amplitude and frequency
 in the radio frequency range provides an efficient solution to assist the magnetization dynamics. Mathematically,
the problem is somewhat similar to but different from the adiabatic population transfer in atomic two-level systems \cite{Cai,Klughertz}, since the magnetic anisotropy is present.
Taking into account the thermal fluctuation or damping parameters, several works pursue the optimal microwave fields
for achieve magnetization dynamics, particularly, the magnetization reversal \cite{Barros11} and switching \cite{Barros13}. But the numerical calculations are sometimes costly,
and the optimally chirped microwave fields are cumbersome for practical implementation. In addition,
the nonlinear spin dynamics and chaotic behavior induced by magnetic anisotropy or competition between damping and pumping
make the magnetization reversal complicated and even disaster \cite{Bertotti,BertottiJAP}.

The main purpose of this paper is to combine the inverse engineering and composite pulses for achieving fast and robust magnetization reversal in ferromagnetic nanostructures.
The magnetization dynamics of a single-domain uniaxial magnetic particle is described by the Landau-Lifshitz-Gilber (LLG) equation, in a circularly polarized ac field of constant amplitude but chirped
frequency. We first apply the inverse engineering method to design the variable frequency for a given short time and amplitude of the ac field, by assuming the prefect quantum two-level system,
and construct the composite pulse to suppress the nonlinear effect resulting from the magnetic anisotropy. But the robustness will be affected by damping parameters, especially when
increasing the composite sequence.  Finally, the rapid magnetization reversal with high fidelity
has been demonstrated with numerical examples and visualized on the Bloch sphere.

\section{Model and Hamilton}

We begin with following Hamiltonian, describing the dynamics of a single-domain magnetic particle with uniaxial anisotropy in a circularly polarized ac field,
 \begin{equation}
\mathcal{H}= -K V M_{z}^{2}-V M_{x} h \cos \Phi(t)-V M_{y} h \sin \Phi(t),
\end{equation}
where $K$ is the magnetic anisotropy constant, $V$ is the volume of particles, $\mathbf{M}$ is the magnetization, $h$ is the amplitude of the ac field, and $\Phi(t)$ is the phase, producing
the time-dependent instantaneous frequency $\omega(t)\equiv \dot{\Phi}(t)$. When the frequency is linearly changing with time, the model resembles the Landau-Zener scheme in conventional quantum
(nonlinear) two-level system \cite{Cai}. Here we shall design inversely the nonlinear time-dependent frequency produced by chirped microwave fields for fast magnetization reversal.

In macrospin approximation, by magnetic moment $\mathbf{M}$, with $|M| = \mu_s$, the magnetization dynamics is equivalently described by LLG equation,
\beq
\label{LLG}
\dot{\mathbf{s}} = \gamma\mathbf{ s} \times \mathbf{H} - \alpha \gamma \mathbf{s} \times (\mathbf{s} \times \mathbf{H}),
\eeq
where $\alpha$ is the dimensionless damping coefficient, $\gamma$ is the gyromagnetic ratio, and the effective Hamiltonian, $\mathbf{H} = -(1/V) \partial \mathcal{H}/ \partial \mathbf{M}$, in the rotating frame
\beq
\mathbf{H} =2 d s_z \mathbf{e}_z + h \mathbf{e}_x + \omega(t) \mathbf{e}_z,
\eeq
with the anisotropy field $d= KM_s$. The initial spin state is antiparticle to $\mathbf{e}_z$ axis which can be prepared by
static magnetic field, $B_0$, which can be switched off after the initialization. The typical experimental parameters for 3-nm-diameter cobalt nanoparticles are chosen
as: gyromagnetic ratio $\gamma=1.76 \times 10^{11} {TS}^{-1}$, anisotropy constant $K=2.2 \times 10^5~J/m^3 $, a volume $V=14.1 \times 10^{-27}~m^{3}$,
$M_s = 1.44 \times 10^{6}~A m^{-1}$, and magnetization at saturation $\mu_s =2.36 \times 10^{-20}~J/T$ \cite{Barros13,Klughertz}. For the sake of
simplicity, we introduce the normalized field $h\equiv h/h_0$ with $h_0 = 2K/\mu_0 M_s \simeq 305~mT$, and the dimensionless time corresponds to $t/t_0$ with $t_0= 1/(\gamma h_0)\simeq 1.86 \times 10^{-11}~s$.

To adopt the inverse engineering, proposed in Ref. \cite{Ruschhaupt,YuePRL}, we first consider the dissipationless problem, when $\alpha=0$.
The LLG equation in general can be parameterized by the convenient spherical coordinates,
\beq
s_z= \cos \theta,~~ s_x= \sin \theta \cos \varphi,~~ s_y= \sin \theta \sin \varphi,
\eeq
and we obtain, by neglecting the ac field in the dissipation terms, as
\beqa
\label{inversetheta}
\dot{\theta} &=& \gamma h \sin \varphi - \alpha \gamma d \sin 2 \theta,
\\
\label{inversevarphi}
\dot{\varphi } &=& -2\gamma d \cos \theta -\omega (t)+\gamma h \cos \varphi \cot \theta.
\eeqa
Actually, in presence of magnetic anisotropy, the spin system resembles the nonlinear two-level systems in BEC in double-well potential \cite{ChenYA} and accelerated optical lattices \cite{Dou,DouComposite}
and coupled waveguides \cite{Khomeriki}. One can apply the Eqs. (\ref{inversetheta}) and (\ref{inversevarphi}) to engineer inversely the amplitude and frequency of microwave fields,
and the time-optimal solution has been obtained accordingly \cite{ChenPRA}. Nevertheless, stability might be spoiled due to nonlinearity resulting from magnetic anisotropy.
Moreover, by assuming that magnetic anisotropic term $d$ is negligible, $d=0$, the above equations
are further simplified as
\beqa
\label{inversetheta2}
\dot{\theta} &=& \gamma h \sin \varphi,
\\
\label{inversevarphi2}
\dot{\varphi} &=& -\omega (t)+\gamma h \cos \varphi \cot \theta.
\eeqa
These are nothing but the auxiliary differential equations, describing the dynamics of population transfer in atomic
two-level systems interacting with laser, based on Lewis-Riesenfeld invariant \cite{Ruschhaupt} or inverse engineering \cite{SarmaPRL,Shore}. Here we shall combine the inverse engineering and
composite pulses, similar to the hybrid method used in Ref. \cite{DouComposite}. Our strategy is to apply
inverse engineering for linear two-level system to design the frequency of microwave fields for fast magnetization reversal,
and further construct the composite pulses by choosing appropriate sequence and phase. This makes the protocol, not only fast but also stable with respect to
the variations of experimental parameters, and magnetic anisotropy.

\begin{figure}[tbp]
\scalebox{0.5}[0.5]{\includegraphics{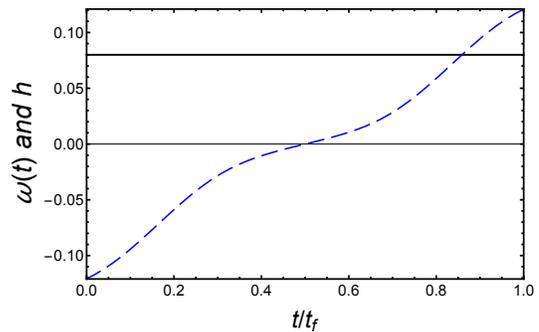}}
\caption{(Color online) Time-dependent frequency of designed microwave fields versus $t/t_0$, where $h=0.08$ is constant and $t_f=100t_0$. }
\label{fig1}
\end{figure}

\section{Inverse engineering and composite pulses}

First of all, we shall design the single pulse of microwave field for speeding up the conventional slowly adiabatic the magnetization reversal by using the inverse engineering method. In principle,
the amplitude and frequency of the ac field are both time-dependent and tunable. But the variation of amplitude could be complicated for physical implementation,
and could induce the amplitude noise. For simplicity, we consider the Landau-Zener type scheme, that is, the constant amplitude but variant (chirped) frequency.
To this end, we rewrite the Eq. (\ref{inversetheta2}) by taking the second derivative with respect to time, and obtain
\beq
\ddot{\theta} = \dot{\varphi} \gamma h \cos \varphi,
\eeq
from which, by combining Eq. (\ref{inversetheta2}) and substituting into Eq. (\ref{inversevarphi2}), we have
\begin{equation}
\label{frequency}
\omega (t) =- \frac{\ddot{\theta }}{\gamma h \sqrt{1-\left( \frac{\dot{\theta }}{\gamma h} \right)^{2}} }+\gamma h \cot \theta\sqrt{1-\left( \frac{\dot{\theta }}{\gamma h} \right)^{2}}.
\end{equation}
This gives the chance to engineer the chirped frequency when the spin trajectory is designed first. But from Eq. (\ref{frequency}) the condition,
$\dot \theta  \leq \gamma h$, which implies the operation time $t_f$ should satisfy, $t_f \geq \pi/\gamma h ~(\approx 40 t_0)$, for magnetization reversal.
Noting that the minimum time for Landau-Zener type scheme is $\pi/\gamma h$, corresponding to constant $\pi$ pulse, and it can be also
achieved for unconstrained driving in our system, when $\omega(t) = d \cos \theta$ for cancelling the nonlinearity \cite{ChenPRA}.

Now, we use the inverse engineering method by choosing the following boundary conditions,
\beqa
\theta(0) &=& \pi, ~~~~~~ \theta (t_f) = 0, \\
\dot{\theta} (0) &=& -\gamma h, ~~ \dot{\theta} (t_f) = -\gamma h, \\
\ddot{\theta} (0) &=& 0, ~~~~~~ \ddot{\theta} (t_f) = 0.
\eeqa
The first two conditions guarantee the magnetization reversal, and others make the frequency smooth at the edges and without singularity.
The simple polynomial ansatz $\theta = \sum_0^j a_j t^j$, where the coefficients $a_j$ are analytically solved from boundary conditions.
Once $\theta$ is interpolated, $\varphi$ is determined by $\varphi= \sin^{-1} (\theta/\gamma h)$. The chirped frequency is finally
designed, from Eq. (\ref{frequency}), see Fig. \ref{fig1}, where $t_f= 100 t_0 $. The advantage of inverse engineering
is that the chirped pulse can be designed for a given constant amplitude within short time, as compared to
adiabatic control, with Landau-Zener scheme. The shortcut design with only $\sigma_z$ control is also different from
the previous ones presented in Ref. \cite{Ruschhaupt} where both Rabi frequency and detuning, corresponding to the amplitude and frequency
of field, can be modulated simultaneously.

\begin{figure}[tbp]
\scalebox{0.8}[0.8]{\includegraphics{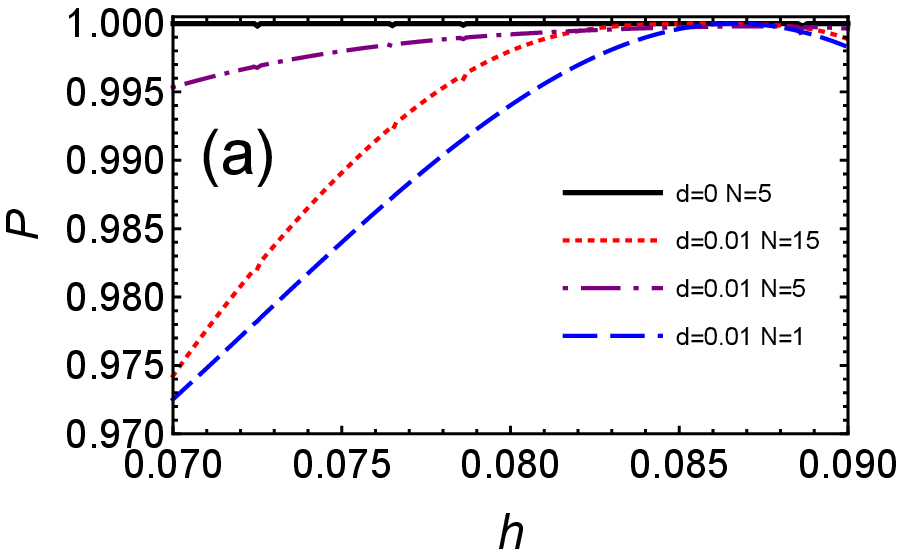}}
\scalebox{0.8}[0.8]{\includegraphics{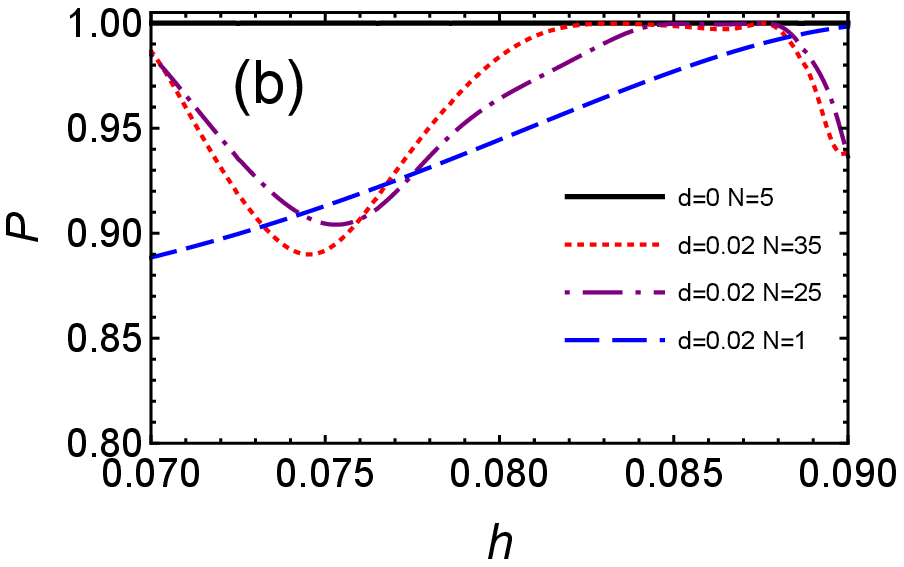}}
\caption{(Color online) Probability of spin-up state at the final time $t=t_f$ versus the amplitude, $h$, of microwave fields, with different magnetic anisotropy $d$ and composite sequence $N$.
The parameters are the same as those in Fig. \ref{fig1}. }
\label{fig2}
\end{figure}

Next, the composite pulse can reduce significantly the error and suppress the nonlinear effect, even with simple
three- and five-pulse composite sequences. Keeping this in mind, we construct the composite pulses with
a sequence of $N$ ($N = 2n + 1$, $n$ is an integer) pulses, each with a phase $\phi_k$ ($k = 1,2,...N$), to achieve high-fidelity
quantum control. The phase $\phi_k$ is imposed on the amplitude of ac field, $h \rightarrow h e^{i \phi_k}$. To shorten the total operation time,
we first try shortcut to adiabatic protocol presented above for nonlinear system, the composite control phase in the linear
systems is exploited here. In detail, the composite phase is given by \cite{Torosov-PRL,DouComposite}
\begin{equation}
\phi_k= \left(N+1-2\left[ \frac{k+1}{2}\right] \right)\left[ \frac{k}{2} \right]\frac{\pi}{N},
\end{equation}
where the symbol $[x]$ denotes the floor function. The phase sequence is symmetric, i.e., $\phi_k = \phi_{N+1-k}$ and $\phi_1=\phi_N=0$.

\begin{figure}[tbp]
\scalebox{0.8}[0.8]{\includegraphics{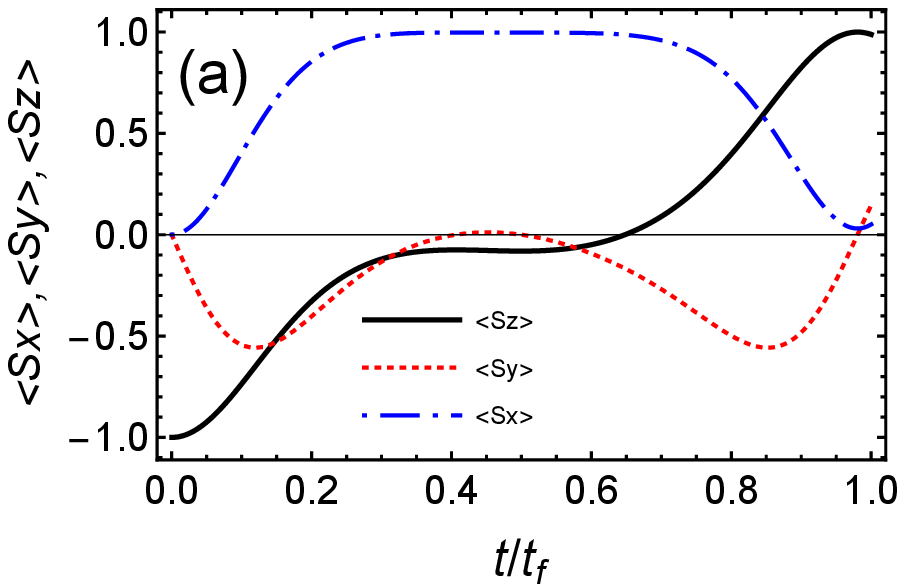}}
\scalebox{0.8}[0.8]{\includegraphics{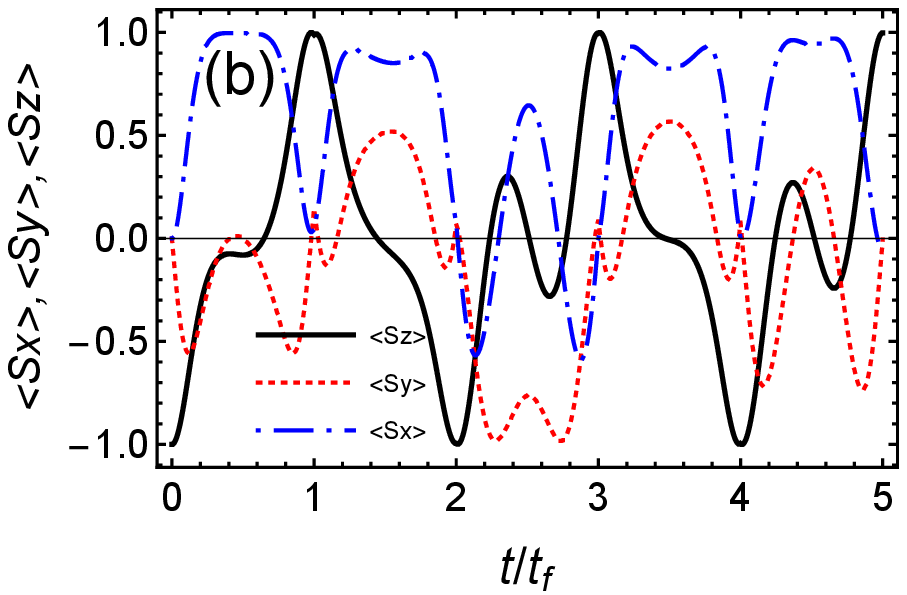}}
\scalebox{0.45}[0.45]{\includegraphics{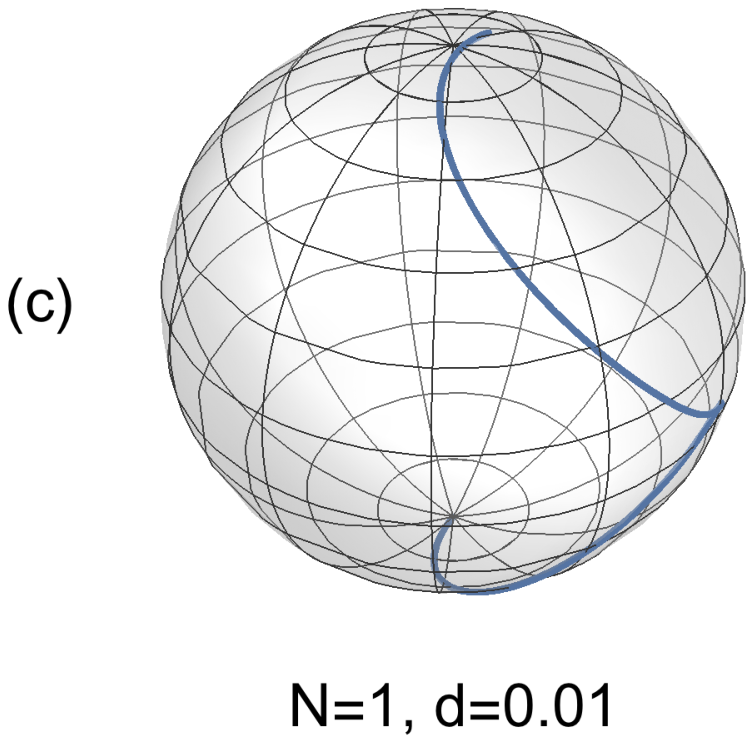}}
\scalebox{0.45}[0.45]{\includegraphics{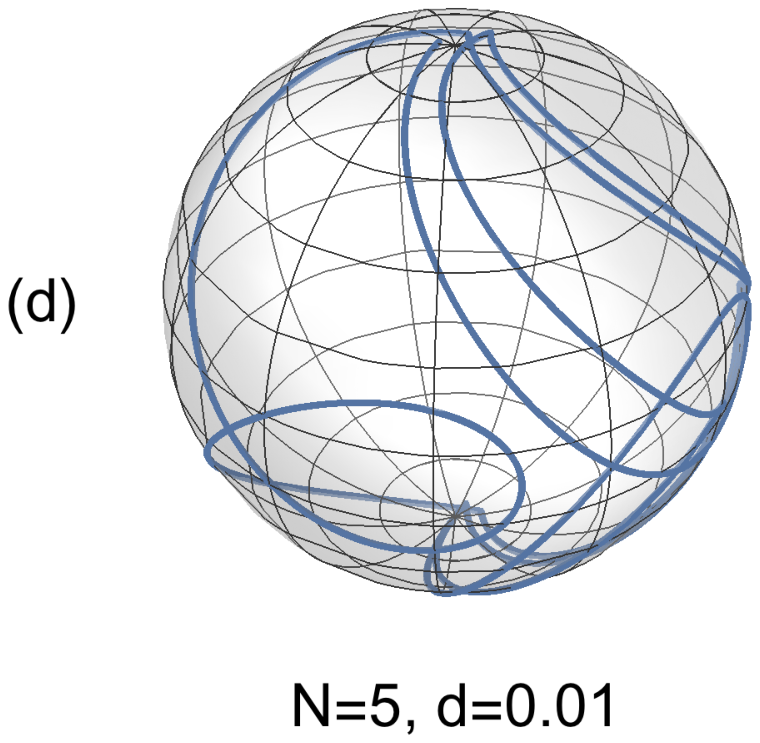}}
\caption{(Color online) Magnetization dynamics $\langle s_j \rangle$ ($j=x,y,z$) and trajectory on Bloch sphere, where (a,c) $N=1$ and (b,d) $N=5$, other parameters are $d=0.01$, $h=0.08$, and $t_f=100 t_0$ for each pulse. }
\label{fig3}
\end{figure}

\section{Examples and Discussions}

We are concerned about the magnetization reversal, and thus the probability $P$ of spin-up state at the final time $t_f$, parallel to $\mathbf{e}_z$ direction.
Fig. \ref{fig2} demonstrates that the perfect magnetization reversal can be achieved by single or composite pulses, when the magnetic anisotropy
$d$ is not involved. However, when magnetic anisotropy, $d$, increasing, the probability is reduced dramatically, and the spin cannot be completely flipped even for large $d$.
Furthermore, the composite pulses even with five composite sequences can improve the probability, taking into account the magnetic anisotropy.
For a certain amplitude, $h$, of the microwave field, the magnetization reversal can be perfect, that is, the probability of spin-up state is almost $1$.
This result coincides with the nonlinear two-level system \cite{DouComposite}, implemented in accelerated optical lattice, in which the composite pulse can suppress the nonlinear effect.

\begin{figure}[tbp]
\scalebox{0.7}[0.7]{\includegraphics{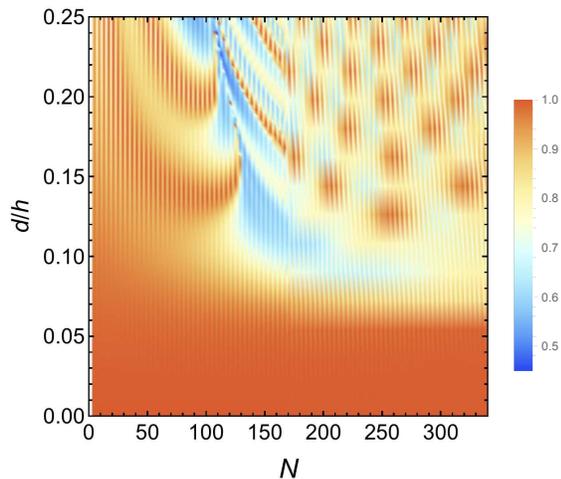}}
\caption{(Color online) Counterplot of probability of spin-up state at the final time $t=t_f$ versus composite sequence $N$ and magnetic anisotropy $d$, where $h=0.05$ and $t_f =100 t_0$. }
\label{fig4}
\end{figure}

\begin{figure}[]
\scalebox{0.8}[0.8]{\includegraphics{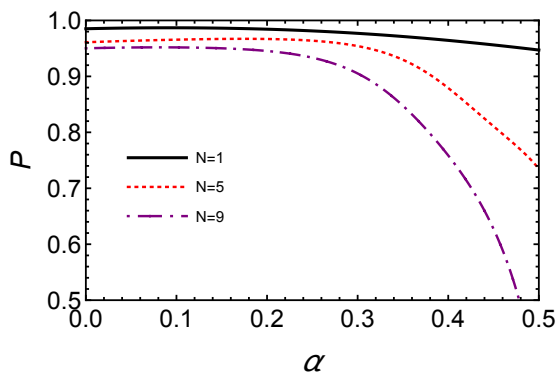}}
\caption{(Color online) Probability at the final time $t=t_f$ versus the damping parameter $\alpha$ for different composite sequence $N$, where $t_f=100 t_0$, $h=0.05$ and $d=0.005$.}
\label{fig5}
\end{figure}

In presence of magnetic anisotropy, $d \neq 0$, the magnetization dynamics cannot be described by the parameters $\theta$ and $\varphi$, since the amplitude and frequency are inversely engineered
from Eqs. (\ref{inversetheta2}) and (\ref{inversevarphi2}), in the context of linear two-level systems. Fig. \ref{fig3} shows the comparison between the cases of a single pulse and five composite sequences.
When $d=0.01$, the magnetization reversal is not perfect by using single pulse, $N=1$, see Fig. \ref{fig3} (a).
Remarkably, the composite pulse even with five sequences works well, see Fig. \ref{fig3} (b). The corresponding trajectory of magnetization dynamics
are both shown in Fig. \ref{fig3} (c) and (d). Anyway, by increasing $N=1$ to $N=5$, the probability is improved from $0.994$ to $0.999$. One
can also choose large composite sequences, but the probability is not always increased, see the discussion below.

Figure \ref{fig4} displays the final probability as a function of composite sequence $N$ and magnetic anisotropy $d$. When the influence of magnetic anisotropy is negligible, the magnetization reversal can be achieved
for single and composite pulses. However, the composite pulses with more sequences are required to improve the stability and suppress the nonlinear effect, when magnetic anisotropy increasing. More interestingly,
we see from Fig. \ref{fig4} that the probability at the final time $t=t_f$ oscillates with composite sequence $N$ and magnetic anisotropy $d$. We identify that
the hybrid method combining the inverse engineering and composite pulses has some advantage over the acceleration and stability.

Finally, we turn to discuss the influence of damping parameter $\alpha$, described in LLG equation (\ref{LLG}). The probability at the final time $t=t_f$ for each composite pulses is
reduced when increasing the damping parameters. In addition, for a larger composite sequence $N$, the longer operation time makes the final probability less. In this sense, the damping
parameter works as the dephasing noise. One can use the technique of shortcuts to adiabaticity to decrease the time for each pulse, thus avoiding the influence of damping. But the ability of
shortening time is in principle limited by the amplitude of microwave fields. So one has to keep the balance by choosing appropriate pulse shape, duration and composite sequence for a given $h$ and $d$.

\section{Conclusion}
In summary, the method for achieving fast and robust magnetization reversal is proposed in a nanomagnet by combing the inverse engineering composite pulses.
The inverse engineering is first applied to design a fast magnetization reversal for each pulse with time-dependent chirped frequency. To suppress the
effect of magnetic anisotropy and improve the stability, the composite pulses are further incorporated later. The magnetization reversal with high
fidelity have been demonstrated with numerical examples. The influence of field amplitude, magnetic anisotropy and damping parameters are also discussed showing
the advantage of hybrid methods. There are several works for further exploration, for example, the optimization of pulses with respect to different errors and noise \cite{Ruschhaupt} and
the effect of thermal fluctuations \cite{Klughertz}. Besides, alternative approaches, based on Lewis-Riesenfeld invariant, can be tried for
the magnetization reversal in the non-equilibrium domain \cite{Ho,Saradzhev}. Last but not least, we hope the fast and robust magnetization reversal or switching can be applicable in high-density information storage and processing in ferromagnetic nanostructures.

\section*{ACKNOWLEDGMENTS}
This work is partially supported by the NSFC (11474193, 61404079), the Shuguang and Yangfan Program (14SG35, 14YF1408400), the Specialized Research Fund for the Doctoral Program (2013310811003),
and the Program for Professor of Special Appointment (Eastern Scholar).

\end{document}